\documentclass[twocolumn,prl,showpacs,amsmath,showkeys,amssymb]{revtex4}

\usepackage{graphicx}
\usepackage{dcolumn}
\usepackage{bm}

\begin{document}


\title{Long-range ferromagnetic dipolar ordering of high-spin molecular clusters}
\author{A. Morello $^{1}$, F.L. Mettes $^{1}$, F. Luis $^{2}$, J. F. Fern\'andez $^{2}$,
J. Krzystek $^{3}$, G. Arom\'{\i} $^{4}$, G. Christou $^{5}$, and
L.J. de Jongh $^{1}$}

\affiliation{$^1$ Kamerlingh Onnes Laboratory, Leiden Institute of Physics, Leiden University, P.O. Box 9504, 2300 RA Leiden, The Netherlands.\\
$^{2} $Instituto de Ciencia de Materiales de Arag\'on, CSIC-Universidad de Zaragoza, 50009 Zaragoza, Spain.\\
$^{3} $National High Magnetic Field Laboratory, Tallahassee, FL 32310, USA.\\
$^{4} $Gorlaeus Laboratories, Leiden Institute of Chemistry, Leiden University, P.O. Box 9502, 2300 RA Leiden, The Netherlands.\\
$^{5} $Department of Chemistry, University of Florida,
Gainesville, FL 32611, USA.\\}
\date{\today}

\begin{abstract}
We report the first example of a transition to long-range magnetic
order in a purely dipolarly interacting molecular magnet. For the
magnetic cluster compound
Mn$_{6}$O$_{4}$Br$_{4}$(Et$_{2}$dbm)$_{6}$, the anisotropy
experienced by the total spin $S=12$ of each cluster is so small
that spin-lattice relaxation remains fast down to the lowest
temperatures, thus enabling dipolar order to occur within
experimental times at $T_c = 0.16$ K. In high magnetic fields, the
relaxation rate becomes drastically reduced and the interplay
between nuclear- and electron-spin lattice relaxation is revealed.
\end{abstract}

\pacs{75.10.Jm, 75.30.Kz, 75.45.+j}

\maketitle

Few examples of long-range magnetic order induced by purely
dipolar interactions are known as yet \cite{White93,Bitko96}.
Therefore, the possibility to study such phase transitions and the
associated long-time relaxation phenomena in detail in high-spin
molecular cluster compounds, with varying crystalline packing
symmetries and different types of anisotropy, presents an
attractive subject \cite{Fernandez00-02}. However, for the most
extensively studied molecular clusters sofar such as Mn$_{12}$,
Fe$_8$ and Mn$_4$ \cite{Sangregorio97,Aubin98,Thomas99,Mettes01}
the uniaxial anisotropy experienced by the cluster spins is very
strong. Consequently, the electronic spin-lattice relaxation time
$T_{1}^{el}$ becomes very long at low temperatures and the cluster
spins become frozen at temperatures of the order of 1 K, i.e. much
higher than the ordering temperatures $T_c \sim 0.1$ K expected on
basis of the intercluster dipolar couplings \cite{Fernandez00-02}.
Although quantum tunneling of these cluster spins has been
observed \cite{Sangregorio97,Aubin98,Thomas99,Mettes01}, and could
in principle provide a relaxation path towards the magnetically
ordered equilibrium state \cite{Fernandez00-02}, the associated
rates in zero field are extremely small ($<100$ Hz). For these
systems, tunneling only becomes effective when strong transverse
fields $B_{t}$ are applied to increase the tunneling rate.
Although the tunability of this rate and thus of $T_{1}^{el}$ by
$B_{t}$ could recently be demonstrated for Mn$_{12}$, Fe$_8$ and
Mn$_4$ \cite{Mettes01}, no ordering has yet been observed.

The obvious way to obtain a dipolar molecular magnet is thus to
look for a high-spin molecule having sufficiently weak magnetic
anisotropy and negligible {\em inter}-cluster super-exchange
interactions. Here we report data for
Mn$_{6}$O$_{4}$Br$_{4}$(Et$_{2}$dbm)$_{6}$, hereafter abbreviated
as Mn$_{6}$ \cite{Aromi99}. The molecular core of Mn$_{6}$ is a
highly symmetric octahedron of Mn$^{3+}$ ions (with spin $s=2$)
that are ferromagnetically coupled via strong {\em intra}-cluster
super-exchange interactions (Fig. 1). Accordingly, the ground
state is a $S=12$ multiplet and the energy of the nearest excited
state is approximately $150$ K higher \cite{Aromi99}. The unit
cell is monoclinic, with space group Pc and contains $4$ molecules
\cite{crystal} bound together only by Van der Waals forces. {\em
Inter}-cluster superexchange is therefore negligible and only
dipolar interactions couple the cluster spins. The net
magnetocrystalline anisotropy of this cluster proves to be
sufficiently small to enable measurements of its equilibrium
magnetic susceptibility and specific heat down to our lowest
temperatures ($15$ mK).

\begin{figure}[b]
\includegraphics[width=8cm]{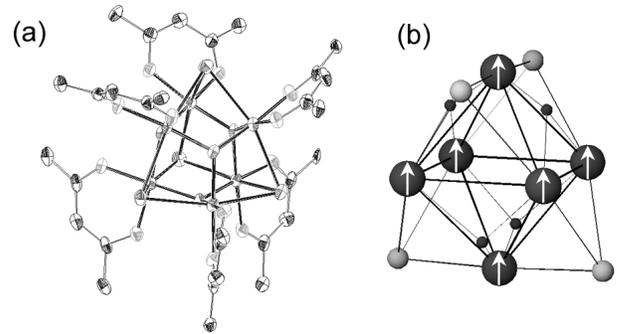}
\caption{\label{fig1} (a) ORTEP representation \cite{Aromi99} of
the structure of Mn$_{6}$O$_{4}$Br$_{4}$(Et$_{2}$dbm)$_{6}$; (b)
detail of the symmetric octahedral core, containing six
ferromagnetically coupled Mn$^{3+}$ ions, yielding a total spin $S
= 12$ for this molecular superparamagnetic particle.}
\end{figure}

The main results of this paper follow. We have observed that the magnetic clusters do undergo a transition to a
long-range ferromagnetically ordered state at $T_{c}=0.161(2)$ K. This transition can be observed within
experimental times ($1-100$ s) because the magnetic relaxation is not too slow at very low temperatures. We have
obtained further information on the magnetic relaxation by applying a magnetic field. It turns out that the
relaxation rate vanishes exponentially as the field increases. The interplay between nuclear- and electron-spin
lattice relaxation is also revealed.

Polycrystalline samples of Mn$_{6}$ were prepared as in Ref.
\cite{Aromi99}. The specific heat of a few milligrams of sample,
mixed with Apiezon grease, was measured at low-T in a home-made
calorimeter \cite{Mettes01} that makes use of the thermal
relaxation method. An important advantage of this method is that
the characteristic time $\tau_{e}$ of the experiment (typically,
$\tau_{e} \simeq 1 - 100$ seconds at low-T) can be varied by
changing the dimensions (and therefore the thermal resistance) of
the Au wire that acts as a thermal link between the calorimeter
and the mixing chamber of the $^3$He-$^4$He dilution refrigerator.
The ac susceptibility data were taken between $0.015$ and $4$ K
with a mutual inductance bridge. The frequency was varied from
$230$ to $7700$ Hz. Magnetic data for $T>1.8$ K were taken with a
SQUID magnetometer.

We first discuss the magnitude of the anisotropy. The spin
Hamiltonian for the molecule is:
\begin{equation}
{\cal H} = -DS_{z}^{2} - g\mu_{B}\vec{B}\cdot\vec{S}.
\label{Hamiltonian}
\end{equation}

\noindent Previously published magnetic data for $T>1.8$ K already
indicated an upper limit of about 0.01 K for $D/k_{B}$
\cite{Aromi99}. In order to obtain an independent estimate of $D$,
high-frequency ESR data were taken in the range $95 - 380$ GHz.
Owing to the combination of very small $D$ and large $S$, as well
as the presence of a signal at $g = 2.00$ arising from a minute
amount of Mn$^{2+}$ impurity (often seen in ESR of Mn$^{3+}$
compounds), the interpretation of the spectra was not fully
conclusive. Nevertheless, signals with a clearly visible structure
on the low-field end of the spectra could be obtained. It could be
identified as fine structure originating from zero-field splitting
(ZFS), since it was independent of field and frequency.
Simulations of the spectra performed using Eq. (\ref{Hamiltonian})
agree well with the experiment taking $|D|/k_{B}\sim 0.03$ or
$\sim 0.05$ K, depending on the sign of $D$ (which could not be
unequivocally determined). Although a smaller rhombic component
could be present, the data do not justify a more elaborate
fitting.

The isotropic character of the molecular spin might seem
paradoxical at first, considering that the individual Mn$^{3+}$
ions, being Jahn-Teller ions, experience strong anisotropy.
However, the spin Hamiltonian for the cluster is determined by the
vectorial addition of the local tensors of the individual atoms,
which can give rise to a low net anisotropy for highly symmetric
molecules such as Mn$_{6}$ (cf. Fig. 1), no matter how large the
ZFS of the constituting atoms \cite{Bencini90}. In fact, the
possibility to tune the net anisotropy of the cluster spin is one
of the attractive properties of molecular superparamagnets.

\begin{figure}[t]
\includegraphics[width=8cm]{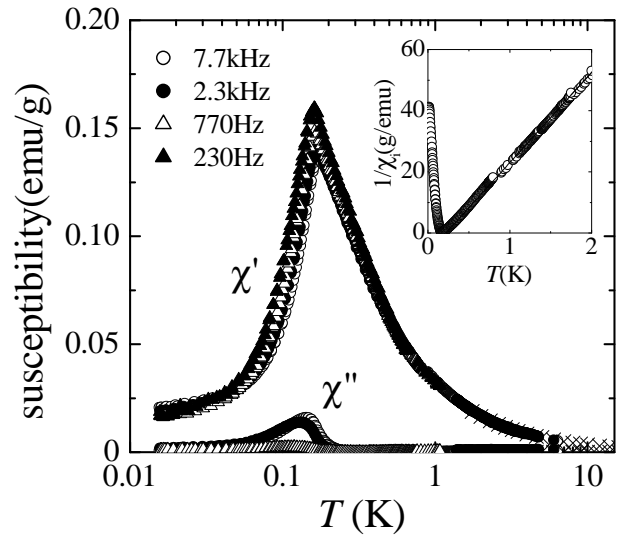}
\caption{\label{fig2} Ac susceptibility of Mn$_{6}$ at zero field
and four different frequencies. Data measured above $1.8$ K in a
commercial SQUID magnetometer are also shown ($\times$). Inset:
$\chi^{\prime}$ corrected for demagnetization effects. The full
line shows the best fit of a Curie-Weiss law to the data for $T
> 0.3$ K, yielding $\theta = 0.20(3)$ K.}
\end{figure}

As first evidence for ordering of the magnetic moments we show
zero-field ac susceptibility data in Fig. 2. The real part
$\chi^{\prime}$ shows a sharp maximum at $T_{c} = 0.161(2)$ K. We
found that $\chi^{\prime}$ at $T_{c}$ is close to the estimated
limit for a ferromagnetic powder sample, $1/(\rho N_{s}+\rho_{sam}
N_{c}) \simeq 0.14 \pm 0.02$ emu/g, where $\rho = 1.45$ g/cm$^{3}$
and $\rho_{sam} \simeq 0.45$ g/cm$^{3}$ are respectively the
densities of bulk Mn$_{6}$ and of the powder sample, $N_{s}=4
\pi/3$ is the demagnetizing factor of a crystallite, approximated
by a sphere, and $N_{c}\simeq 2.51$ is the demagnetizing factor of
the sample holder. This indicates that Mn$_{6}$ is
ferromagnetically ordered below $T_{c}$. Susceptibility data
$\chi_{i}$ corrected for the demagnetizing field ($\chi_{i} =
\chi^{\prime}/[1 - (\rho N_{s} + \rho_{sam} N_{c})
\chi^{\prime}]$) follow the Curie-Weiss law $\chi = C/(T -
\theta)$ down to approximately $0.3$ K, with $C = 0.034(1)$
cm$^{3}$K/g and $\theta= 0.20(3)$ K. The constant $C$ equals,
within the experimental errors, the theoretical value for randomly
oriented crystals with Ising-like anisotropy
$N_{A}g^{2}\mu_{B}^{2}S(S+1)/3k_{B}P_{m} = 0.0332$ cm$^{3}$K/g,
where $S = 12$, $g = 2$, and the molecular weight $P_{m} =
2347.06$. The positive $\theta$ confirms the ferromagnetic nature
of the ordered phase. From the mean-field equation $\theta =
2zJ_{eff}S(S+1)/3k_{B}$, we estimate the effective inter-cluster
magnetic interaction $J_{eff} \approx 1.6 \times 10^{-4}$ K, and
the associated effective field $H_{eff} = 2zJ_{eff}S/g\mu_{B} =
3.5 \times 10^{2}$ Oe coming from the $z=12$ nearest neighbors.

The maximum value of $\chi^{\prime}$ is seen to vary only weakly
with $\omega$, which we attribute to the anisotropy. The total
activation energy of Mn$_{6}$ amounts to $DS^{2} \approx 1.5$ K,
i.e., it is about $45$ times smaller than for Mn$_{12}$.
Accordingly, one expects the superparamagnetic blocking of the
Mn$_{6}$ spins to occur when $T \simeq T_{B}($Mn$_{12})/45$, that
is below $\simeq 0.12$ K. In other words, for $T \rightarrow T_c$,
the approach to equilibrium begins to be hindered by the
anisotropy of the individual molecular spins. We stress however,
that the frequency dependence of $\chi^{\prime}$ observed here is
very different from that of the well known anisotropic
superparamagnetic clusters or that of spin glasses. Below $T_{c}$,
$\chi^{\prime}$ decreases rapidly, as expected for an anisotropic
ferromagnet in which the domain-wall motions become progressively
pinned. The associated domain-wall losses should then lead to a
frequency dependent maximum around $T_c$ in the imaginary part,
$\chi^{\prime \prime}$, as seen experimentally. Indeed, although
the Mn$_6$ spins can be considered as nearly isotropic at high
temperatures, the anisotropy energy ($\simeq 2DS$) is of the same
order as the dipolar interaction energy $\mu^2 / r^3 \simeq 0.1$ K
between nearest neighbor molecules. Thus the ordering should be
that of an Ising dipolar ferromagnet.

\begin{figure}[t]
\includegraphics[width=8cm]{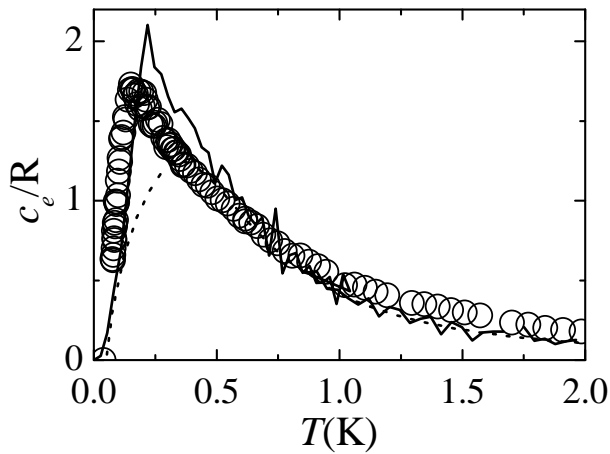}
\caption{\label{fig3} Electronic specific heat data ($\bigcirc$)
of Mn$_6$ at zero field. The dotted line is the Schottky anomaly
calculated with $D=0.013$ K. The full line is the Monte Carlo (MC)
calculation for an orthorhombic lattice of $1024$ Ising spins with
periodic boundary conditions. For each point, we performed $2
\times 10^{4}$ MC steps per spin.}
\end{figure}

Additional evidence for the ferromagnetic transition at $T_{c}$ is
provided by the electronic specific heat $c_{e}$ \cite{Ce}, shown
in Fig. 3, which reveals a sharp peak at $0.15(2)$ K. Numerical
integration of $c_{e}/T$ between $0.08$ K and $4$ K gives a total
entropy change of about $3.4 k_{B}$ per molecule, very close to
the value for a fully-split $S=12$ spin multiplet (i.e., $k_{B}
\ln(2S+1)=3.22k_{B}$). We thus may attribute the peak to the
long-range order of the molecular spins. We note, however, that at
$T_c$ the entropy amounts to about $1 k_B$ per spin, showing that
only the lowest energy spin states take part in the magnetic
ordering.

We have performed Monte-Carlo (MC) simulations for an $S=12$ Ising
model of magnetic dipoles on an orthorhombic lattice with axes
$a_x = 15.7$ \AA, $a_y=23.33$ \AA, and $a_z=16.7$ \AA, which
approximates the crystal structure of Mn$_{6}$. The model includes
dipolar interactions as well as the anisotropy term $-DS_z^2$
given in Eq. (\ref{Hamiltonian}). Neglecting interactions, the ZFS
of the $S=12$ multiplet produced by this crystal field term lead
to a Schottky anomaly in $c_{e}$, as shown by the dotted curve in
Fig. 3. The fit in the range above 0.5 K yields $D/k_B \simeq
0.013$ K, in good agreement with our above estimates. The
intermolecular dipolar interactions remove the remaining
degeneracy of the $|\pm m\rangle$ spin doublets. The MC
simulations show that the ground state is ferromagnetically
ordered, as observed, and predict a shape for $c_{e}$ that is in
very good agreement with the experiment. In Fig. 3, we show
$c_{e}$ calculated assuming all molecular easy ($z$) axes to point
along $a_z$, i.e. one of the two nearly equivalent short axes of
the actual lattice. Similar results were obtained for other
orientations chosen for the anisotropy ($z$) axis. We note that
the Ising simulations give $T_{c} = 0.22$ K, which is slightly
higher than the experimental $T_c=0.161(2)$ K. This difference may
be related to the finite value of the anisotropy. Model
calculations for this crystal structure, assuming classical
anisotropic Heisenberg spins with varying anisotropy
\cite{fernandez03}, show that different ferromagnetic ground
states are possible, depending on the competition between local
crystal field effects and long-range dipolar interactions. The
variation of $T_c$ with anisotropy, as well as the form of the
calculated and observed specific heat anomaly, are specific for
dipolar interactions, and differ widely from the analogues for the
usual superexchange ferromagnets \cite{deJongh01}.

\begin{figure}[t]
\includegraphics[width=8cm]{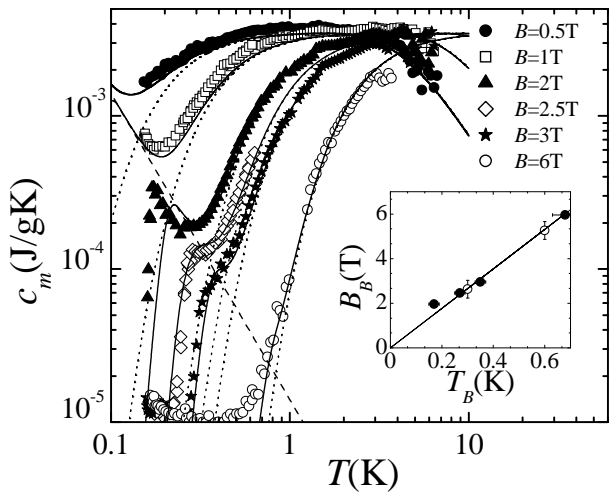}
\caption{\label{fig4} Magnetic specific heat of Mn$_{6}$ for
various applied fields. Dotted and dashed lines: calculated
equilibrium specific heats of the electronic and the nuclear
spins, respectively. Full lines: time-dependent total specific
heat calculated taking into account the nuclear spin-lattice
relaxation. Inset: magnetic field $B_{B}$ needed to take the
nuclear spins off-equilibrium. $\circ$, obtained from $c_{nucl}$
vs $B$ isotherms; $\bullet$, from $c_{nucl}$ vs $T$ at constant
field.}
\end{figure}

We next turn to the specific heat data obtained in varying
magnetic field $B$, plotted in Fig. 4. Even for the lowest $B$
value, the ordering anomaly is fully suppressed, as expected for a
ferromagnet \cite{deJongh01}. Accordingly, we may account for
these data with the Hamiltonian (1) neglecting dipolar
interactions. The Zeeman term splits the otherwise degenerate $|
\pm m \rangle$ doublets, and already for $B \sim 0.5$ T the level
splittings become predominantly determined by $B$, so that the
anisotropy term can also be neglected. As seen in Fig. 4, the
calculations performed with $D=0$ reproduce the data quite
satisfactorily at higher temperatures (dotted curves).

However, when the maxima of the Schottky anomalies are shifted to
higher $T$ by increasing $B$, an additional contribution at low
$T$ is revealed. It is most clearly visible in the curves for $1$
T$ < B < 2.5$ T, and varies with temperature as $cT^{2}/R = 4
\times 10^{-3}$. We attribute this to the high-temperature tail of
the specific heat contribution $c_{nucl}$ arising from the Mn
nuclear spins ($I=5/2$), whose energy levels are split by the
hyperfine interaction with the Mn$^{3+}$ electronic spins $s$.
This interaction can be approximated by ${\cal
H}_{hf}=sA_{hf}m_{I}$, where $A_{hf}$ is the hyperfine constant
and $m_{I}$ is the projection of the nuclear spin along the
electronic spin. At high temperatures (i.e., when $A_{hf} s \ll
k_{B}T$) $c_{nucl}/R \simeq
\frac{1}{3}A_{hf}^{2}s^{2}I(I+1)T^{-2}$ \cite{Abragam70}. Taking
$A_{hf}=7.6$ mK as used previously to simulate ESR spectra
measured on a Mn$_{4}$ cluster \cite{Zheng96}, we obtain the
dashed line of Fig. 4. This contribution was subtracted from the
zero-field data shown in Fig. 3.

A remarkable feature of the experimental data that is not
reproduced by these calculations is that, at the lowest $T$, the
nuclear specific heat drops abruptly to about $10^{-5}$ J/gK. The
temperature $T_{B}$ where the drop occurs depends on $B$ but also
on the characteristic time constant $\tau_{e}$ of our
(time-dependent) specific heat experiment: the deviation from the
(calculated) equilibrium specific heat is found at a lower $T$
when the system is given more time to relax. We conclude that the
drop indicates that nuclear spins can no longer reach thermal
equilibrium within time $\tau_{e}$. We may write
$c_{nucl}(\tau_{e})=c_{nucl}^{eq}[1-\exp(-\tau_{e}/T_{1})]$,
showing that the transition should occur when the nuclear
spin-lattice relaxation time $T_{1}$ becomes of the order of
$\tau_{e}$. These transitions to non-equilibrium provide therefore
direct information on the temperature and field dependence of
$T_{1}$, which can be related to the fluctuation of the transverse
hyperfine field, as produced by the phonon-induced transitions
between different levels of the electronic spin \cite{Abragam61}.

At low $T$ and high $B$, only the ground and the first excited
states, $m=+12$ and $m=+11$, need be considered. We may write
$1/T_{1} \simeq (1/\tau_{0}) \exp\{-[g\mu_{B}B]/k_{B}T \}$, where
$1/\tau_{0}$ plays the role of an attempt frequency for the
electron spin transitions. Clearly, the nuclear spins can be taken
out of of equilibrium either by decreasing $T$ down to $T_B$ at
constant field (as in Fig. 4) or by increasing $B$ up to a given
value $B_B$ at constant $T$. This is indeed observed
experimentally (not shown). The effect of the field is just to
polarize the electronic spins, which reduces the fluctuations of
the hyperfine field, thus effectively disconnecting nuclear spins
from the lattice. For a given $\tau_e$, $B_B$ must increase
linearly with $T_B$, which is confirmed by the experimental data
plotted in the inset of Fig. 4. The slope gives $\tau_0 \approx
3\times 10^{-4}$ s. Using this value, we have calculated the
time-dependent $c_{nucl}$, shown as the full lines in Fig. 4, and
seen to be in reasonable agreement with the experimental data at
all $T$ and $B$.

The authors have enjoyed illuminating discussions with Dr. E.
Palacios, and Prof. P. C. E. Stamp. This work is part of the
research program of the "Stichting voor Fundamenteel Onderzoek der
Materie" (FOM). F. L. acknowledges a TMR grant from the European
Union. J. F. F. acknowledges grant BMF2000-0622 from DGESIC of
Spain.

\noindent $^*$ To whom all correspondence should be addressed.
E-mail address: dejongh@phys.leidenuniv.nl

\bibliographystyle{unsrt}
 \bibliography{MorelloMn6cond-mat}

\begin{thebibliography}{10}

\bibitem{White93}
{S. J. White \textit{et al.}, Phys. Rev. Lett. {\bf71}, 3553 (1993).}

\bibitem{Bitko96}
{see e.g. D. Bitko, T.F. Rosenbaum, and G. Aeppli, Phys. Rev. Lett. {\bf77},
  940 (1996); G. Mennenga, L. J. de Jongh, and W. J. Huiskamp, J. Magn. Mag.
  Mat. {\bf44}, 59 (1984).}

\bibitem{Fernandez00-02}
{J. F. Fern\'andez and J. J. Alonso, Phys. Rev. B {\bf62}, 53 (2000); J. F.
  Fern\'andez, Phys. Rev. B {\bf66}, 064423 (2002).}

\bibitem{Sangregorio97}
{C. Sangregorio \textit{et al.}, Phys. Rev. Lett. {\bf 78}, 4645 (1997).}

\bibitem{Aubin98}
{S. M. Aubin \textit{et al.}, J. Am. Chem. Soc. {\bf 120}, 4991 (1998).}

\bibitem{Thomas99}
{L. Thomas, A. Caneschi, and B. Barbara, Phys. Rev. Lett. {\bf 83}, 2398
  (1999).}

\bibitem{Mettes01}
{F. L. Mettes \textit{et al.}, Phys. Rev. B {\bf 64}, 174411 (2001); F. Luis
  \textit{et al.}, Phys. Rev. Lett. {\bf 85}, 4377 (2000); F. L. Mettes
  \textit{et al.}, Polyhedron, {\bf 20}, 1459 (2001).}

\bibitem{Aromi99}
{G. Arom\'{\i} \textit{et al.}, J. Am. Chem. Soc. {\bf 121}, 5489 (1999).}

\bibitem{crystal}
{We have used the crystallographic data for a related complex
  Mn$_{6}$O$_{4}$Cl$_{4}$(Et$_{2}$dbm)$_{6}$ \cite{Aromi99}, in which the
  Br$^{-}$ ions are replaced by Cl$^{-}$.}

\bibitem{Bencini90}
{A. Bencini and D. Gatteschi, {\it EPR of Exchange Coupled Systems},
  Springer-Verlag (Berlin and Heidelberg, 1990).}

\bibitem{Ce}
{The electronic specific heat was obtained by subtracting from the total
  specific heat the contribution of the lattice, which follows the well-known
  Debye approximation for low temperatures $c_{latt} \propto
  (T/\Theta_{D})^{3}$, where $\Theta_{D} \simeq 48$ K, as well as the
  contribution $c_{nucl}$ arising from the nuclear spins, as discussed in the
  text.}

\bibitem{fernandez03}
{J. F. Fern\'andez and F. Luis, unpublished.}

\bibitem{deJongh01}
{See e.g., L. J. de Jongh and A. R. Miedema, Adv. Phys. {\bf 50}, 947 (2001).}

\bibitem{Abragam70}
{A. Abragam and A. Bleaney, {\it Electron paramagnetic resonance of transition
  ions} (Clarendon Press, Oxford, 1970).}

\bibitem{Zheng96}
{M. Zheng and G. C. Dismukes, Inorg. Chemistry {\bf 35}, 3307 (1996).}

\bibitem{Abragam61}
{A. Abragam, {\it The principles of nuclear magnetism} (Clarendon Press,
  Oxford, 1961). For a recent application of the theory to Mn$_{12}$, see
  Furukawa {\it et al.}, Phys. Rev. B {\bf 64}, 104401 (2001).}

\end{thebibliography}

\end{document}